\documentclass[letterpaper,11pt,fleqn]{article}
 \pdfoutput=1
\usepackage{jheppub}

\usepackage{dcolumn}

\usepackage{graphicx}
\usepackage{bm,amsmath,amssymb}
\usepackage[mathscr]{eucal}

\newcommand{\beq}{\begin{equation}}
\newcommand{\eeq}{\end{equation}}

\long\def\comment#1{ }

\title{  Holographic Picture of  Heavy Vector Meson Melting  }
 
\author[a]{Nelson R. F. Braga}
\author[b]{, M. A. Martin Contreras}
\author[a]{and Saulo Diles}

\affiliation[a]{Instituto de F\'{\i}sica,
Universidade Federal do Rio de Janeiro, Caixa Postal 68528, RJ
21941-972 -- Brazil}
\affiliation[b]{ High Energy Group, Department of Physics, Universidad de los Andes, Carrera 1, No 18A - 10, Bloque Ip, ZIP 111711, Bogot\'{a}, Colombia}

\emailAdd{braga@if.ufrj.br}
\emailAdd{smdiles@if.ufrj.br}
\emailAdd{ma.martin41@uniandes.edu.co}

\abstract{ The fraction of heavy vector mesons produced in a heavy ion collision, as compared to a proton proton collision, serves as an important indication of the formation of a thermal medium, the quark gluon plasma.  This sort of analysis strongly depends on understanding the thermal effects of a medium like the plasma on the states of heavy mesons.  In particular, it is crucial to know the
temperature ranges where they undergo a thermal dissociation, or melting. 

AdS/QCD models are know to provide an important tool for the calculation of hadronic masses, but in general are not consistent with the observation that decay constants of heavy vector mesons decrease with excitation level.
It has recently been shown that this problem can be overcome using a soft wall background and introducing an  extra energy parameter,  through the calculation of  correlation functions at a finite  position of anti-de Sitter space. This approach leads to the evaluation of masses and decay constants of S wave quarkonium states 
with just one flavor dependent and one flavor independent parameters. 

Here we extend this more realistic model to finite temperatures and analyse the thermal behavior of the states $1S, 2S$ and $ 3S$ of bottomonium and  charmonium. 
The corresponding spectral function exhibits a consistent picture for the melting of the states where, for each flavor,  the higher excitations melt at lower temperatures. We estimate for these six states,  the energy ranges in which the heavy vector mesons undergo a transition from a well defined peak in the spectral function to  complete melting  in the thermal medium.  A very clear distinction between the heavy flavors emerges, with bottomonium state $\Upsilon  (1S)$  surviving deconfinemet transition at temperatures much larger than the critical deconfinement temperature of the medium.  
 
        }


\begin{document}
\maketitle

\section{ Introduction }   
The suggestion  \cite{Matsui:1986dk} (see \cite{Satz:2005hx}  for a review) that $J/\psi $   supression in heavy ion collisions could be  a signature for the formation of  quark gluon plasma gave rise to a continuous interest in
the thermal behavior of charmonium states. In particular,  it is of great interest to know what are the temperature ranges  at which the heavy vector mesons states melt. By melting one means the thermal dissociation in the medium that  corresponds to the disappearance of the particle peak in the spectral function. 

AdS/QCD models are very useful tools for studying spectral properties of hadronic states.  Such models,  inspired in the AdS/CFT correspondence \cite{Maldacena:1997re,Gubser:1998bc,Witten:1998qj},  assume the existence of an approximate duality 
 between a field theory living in 
an anti-de Sitter background deformed by the introduction of a dimension-full parameter and a gauge theory where the parameter plays the role of an energy scale.
One of the earliest formulations, the hard wall  AdS/QCD model,  appeared in refs. \cite{Polchinski:2001tt,BoschiFilho:2002ta,BoschiFilho:2002vd} and consists in placing a hard geometrical cutoff in anti-de Sitter (AdS)  space. In particular,  the hard wall model was used in \cite{BoschiFilho:2002ta,BoschiFilho:2002vd} as a tool for calculating masses of glueballs.  
 Another AdS/QCD model, the soft wall, where  the square of the mass  grow linearly with the
radial excitation number was introduced in ref. \cite{Karch:2006pv}. In this case, the background involves AdS space and a scalar field that acts effectively as a smooth  infrared cutoff.  
A recent recent review of AdS/QCD with a wide list of   references can be found in \cite{Brodsky:2014yha}.

AdS/QCD models provide also a tool for calculating another important property of hadrons: the decay constant. The decay of mesons is represented 
as a transition from the initial  state to the hadronic vacuum.  For a meson  at radial excitation level $n$ with
mass $m_n$ the decay constant $  f_n $ is defined as: $  \langle 0 \vert \, J_\mu (0)  \,  \vert n \rangle = \epsilon_\mu f_n m_n $, where $ J_\mu $ is the gauge current and $ \epsilon_\mu $ the polarization.
Expressing the two point correlator of gauge currents as a sum over transition matrix elements,  one  
 finds a holographic expression for decay constants \cite{Karch:2006pv,Grigoryan:2007my}.

A problem of the original formulations of the hard wall and soft wall models is that  the  experimental results available for charmonium and bottomonium vector states  show that higher excited radial states have smaller decay constants. In other words, $f_n$ decrease with $n$. In contrast, the results obtained for decay constants of vector mesons in the soft wall  are degenerate: all the decay constants of the radial excitations of a vector meson are equal. For the hard wall model the decay constants of radial excitations increase with the excitation level.  A fit of the decay constants of charmonium states in soft wall appeared in ref. \cite{Grigoryan:2010pj}, but introducing three extra parameters in the model. 
In ref. \cite{Grigoryan:2010pj} four experimental data,  the masses and decay constants of $J /\psi $ and $\psi^\prime$,
 are used to fix four parameters introduced in the model, thus the formulation lacks of predictivity
 as a model for heavy vector mesons.

An alternative version of the soft wall model,  consistent with the observed behavior  of decay constants, was recently proposed in ref. \cite{Braga:2015jca}.  In contrast to the original formulation, in this new framework the decay constants are obtained from two point correlators of gauge theory operators  calculated at a  finite value $ z = z_0$  of the radial coordinate of AdS space.   This way an extra energy parameter 
  $ 1/z_0 $, associated with  an ultraviolet (UV)  energy scale is introduced  in the model. The masses and decay constants of charmonium
and bottomonium $S$ wave states are calculated in ref. \cite{Braga:2015jca}  using the quantity 
$ 1/z_0 $ as a flavour independent parameter and taking the usual infrared (IR) soft wall  parameter $k$  to depend on the flavor,  since it is associated with the quark mass.   A total of eight masses and eight decay constants are determined using three parameters. The rms error is of $30 \% $ that is reasonable, given the simplicity of the model and the fact that two different properties of two different flavors are adjusted
with just 3 parameters.     

 The purpose of the present article is to extend the model of ref. \cite{Braga:2015jca} to finite temperature in order to investigate the thermal spectra of $S$ wave states of charmonium and bottomonium.  We will show that the spectral functions present the expected behavior:  at low temperatures, sharp peaks for the lower level excitations, and, as the temperature 
 increases, the peaks spread and decrease in height. The evolution of the spectral function with increasing temperature shows clearly the process of transition from  well defined peaks to the disappearance of the states in the medium, for the states $ 1S$ , $2S$ and $3S$. The melting occurs  at lower temperatures for the higher excitations.

 It is worth mentioning that in  refs. \cite{Hong:2003jm,Kim:2007rt,Fujita:2009wc,Noronha:2009da,Fujita:2009ca,Grigoryan:2010pj,Branz:2010ub,Gutsche:2012ez,
Afonin:2013npa,Hashimoto:2014jua},  heavy vector mesons have been discussed in the context of AdS/QCD models.   However,  the  holographic  picture for  the melting of   $ 1S$ , $2S$ and $3S$  states of  bottomonium and charmonium that  we will show here,  was not presented before in the literature.

 The article is organized as follows:  in section 2 we briefly review the model for heavy vector mesons at zero temperature  presented recently in ref. \cite{Braga:2015jca}. Then is section 3 we build up  a finite temperature version for this model and show how to calculate the corresponding thermal spectral function.  In section 4 we show the results obtained by numerically solving the equations of motion. We analyze the melting of the states of charmonium and bottomonium as the temperature increase and estimate the temperature ranges where the thermal dissociation occurs.    We leave for section 5 some final comments and remarks and present in the appendix more details of the melting of charmonium states. Appendix A shows more details of the temperature dependence of the  thermal spectral functions and appendix B presents an analysis of the high frequency behavior.

\section{ Heavy Vector mesons in the vacuum  }

The holographic model proposed in ref. \cite{Braga:2015jca}  contains two dimensionful parameters. One coming from a soft wall background and the other from a position in AdS space where the gauge theory correlators are calculated.  The model leads to decay constants for 
heavy vector mesons decreasing with the radial excitation level, in agreement with the results obtained from experimental data. 
 
   One considers a  vector field $V_m = (V_\mu,V_z)\,$ ($\mu = 0,1,2,3$)  playing the role of the supergravity dual of  the gauge theory current $ J^\mu = \bar{q}\gamma^\mu q \,$. The field lives in a five dimensional soft wall background governed by the action
\begin{equation}
I \,=\, \int d^4x dz \, \sqrt{-g} \,\, e^{- \Phi (z)  } \, \left\{  - \frac{1}{4 g_5^2} F_{mn} F^{mn}
\,  \right\} \,\,, 
\label{vectorfieldactionzeroptemp}
\end{equation}
\noindent where $F_{mn} = \partial_mV_n - \partial_n V_m$ and $\Phi = k^2z^2   $ is the soft wall background, with the parameter  $k$  playng the role of an IR, or mass, energy scale. The space is a Poincar\'e  AdS chart:
\begin{equation}
 ds^2 \,\,= \,\, \frac{R^2}{z^2}  \,  \Big(  - dt^2 + dz^2  + d\vec{x}\cdot d\vec{x}  \Big)   \,.
 \label{metric1}
\end{equation}

The second input parameter of the model, that is not present in the usual formulation of the soft wall model,  is introduced by calculating the correlators at a finite position $ z = z_0$ instead of taking the boundary to be at  $z=0$. The parameter $1/z_0$ is interpreted as an UV energy scale. A similar approach appeared in ref. \cite{Evans:2006ea} but for light vector mesons. 

One considers the action of eq. (\ref{vectorfieldactionzeroptemp}) to be defined in the region $ z_0 \le z <  \infty $, then the on shell action takes the form
 \begin{equation}
I_{on \, shell }\,=\, - \frac{1}{2 {\tilde g}_5^2}  \, \int d^4x \,\,\Big[ \, \frac{e^{- k^2 z^ 2  }}{z} \eta^{\mu \nu} V_\mu \partial_z V_ \mu \,\Big] 
{\Big \vert}_{_{ \! z \to z_0 }} 
 \,,
\label{onshellactionzerotemp}
\end{equation}
where $ {\tilde g}_5^2 =  g_5^2 /R $ is the relevant dimensionless coupling of the vector field and $\eta^{\mu \nu} $ is the Minkowski metric.

The gauge $V_z = 0 $ is used, so that the boundary values of the other remaining  components of the vector field:  
$ V^0_{\mu}(x) =
\lim_{z\to z_0} V_\mu (x,z)$ are the sources of the correlation functions of the boundary current operator  $  J^\mu (x) \,\, (= \bar{q}\gamma^\mu q (x) \,)\,$.  
That means:
\begin{equation}
 \langle 0 \vert \, J_\mu (x) J_\nu (y) \,  \vert 0 \rangle \, =\, \frac{\delta}{\delta V^{0\mu}(x)} \frac{\delta}{\delta V^{0\nu}(y)}
 \exp  \left( - I_{on shell} \right)\,.
\end{equation}

Working in momentum space in the coordinates $x^ \mu$, or equivalently taking a plane wave solution, 
the field $ V_\mu (p,z) $ can be decomposed for convenience into a source factor times a $z$ dependent factor
 \begin{equation} 
 V_\mu (p,z) \,=\, v (p,z) V^0_\mu ( p ) \,,
 \label{Bulktoboundary}
\end{equation}  
 \noindent where    $ v (p,z)   $ is usually called  bulk to boundary propagator and  satisfies the equation of motion:
\begin{equation}
\partial_z \Big( \frac{ e^{-k^ 2 z^ 2 }} { z}  \partial_z v (p,z) \Big) + \frac{p^ 2 }{z} e^{-k^ 2 z^ 2 }  v (p,z) \,=\, 0   \,.
\label{BulktoboundaryEOM}
\end{equation}
In order that the factor $ V^0_\mu ( p )$, defined in the decomposition of eq. (\ref{Bulktoboundary}),  works as the source
of the correlators of gauge theory currents, calculated at $z = z_0 $, one must impose the boundary condition:  
 \begin{equation} 
 v (p, z=z_0) = 1\,.
 \label{boundarycondition}
\end{equation} 
The solution of eq. (\ref{BulktoboundaryEOM}) is a Tricomi function $U (-p^ 2/ 4k^ 2 , 0, k^2 z^ 2) $. The boundary condition can be trivially satisfied following ref. \cite{Afonin:2011ff, Afonin:2012xq}  and
writing: 
\begin{equation} 
 v (p,z ) \, = \, \frac{ U (-p^ 2/ 4k^ 2 , 0, k^2 z^ 2 ) }{U (-p^ 2/ 4k^ 2 , 0, k^2 z_0^ 2 )}\,.
 \label{bulktoboundary2}
\end{equation} 
 The decay constants appear in the two point function:
 \begin{equation}
\Pi (p^2)  = \sum_{n=1}^\infty \, \frac{f_n^ 2}{(- p^ 2) - m_n^ 2 + i \epsilon} \,.
\label{2point}
\end{equation} 
On the other hand, the two point function is related to the current current correlator
 \begin{equation}
  \left( p^2 \eta_{\mu\nu} -  p_\mu p_\nu   \right) \, \Pi ( p^2 ) 
\, =\, \int d^4x \,\, e^{-ip\cdot x} \langle 0 \vert \, J_\mu (x) J_\nu (0) \,  \vert 0 \rangle  \, , 
\label{correlatorand2pointfunction}
\end{equation}  
 that can be obtained holographically by differentiating the on shell action by the boundary values of the fields, with the result:
 \begin{equation}
  \Pi ( p^2 )  \, =\,   \frac{1}{{\tilde g}_5^ 2 \, (-p^ 2) } \left[  \frac{ e^{ -k^2 z^2 }  \,  v (p,z) \partial_z v (p,z)  }{  \, z  } 
  \right]_{_{ \! z \to z_0 }} \, .  
 \label{hol2point}
\end{equation} 

The expression  (\ref{hol2point}) has simple poles, although it does not have the exact simple pole structure of eq. (\ref{2point}).
But one can associate the coefficients of the approximate expansion near the poles with the decay constant $f_n$ in analogy with the exact expansion shown in eq. (\ref{2point}). This way one finds the masses from the localization of the poles of the two point function and the decay constants from the
corresponding  coefficient. That means, if  $\chi_n$ are the roots of the Tricomi function: 
\begin{equation}
 U (  \chi_n \,  , 0, k^2 z_0^ 2 ) = 0 \,,
 \label{roots}
 \end{equation} 
 \noindent then the holographic vector meson masses are:
 \begin{equation} 
 m^2_n \, =  \, 4 k^ 2  \, \chi_n \,.
 \label{new masses}
\end{equation} 
The decay constants are calculated numerically from the fit to the approximate form of the simple pole of  eq. (\ref{2point}).  That means:
\begin{equation}
f_n^ 2 \,=\, 
\lim_{p^2 \to - m^2_n} \Big( (-p^ 2)  - m_n^ 2 \Big) \, 
\Pi (p^2)   \ 
 \,.
\label{numerical decay}
\end{equation} 

The coupling ${\tilde g}_5 = g_5 /\sqrt{R} $ of the vector field in the AdS bulk is obtained by comparison with QCD (see refs.  \cite{Karch:2006pv,Grigoryan:2007my}) , wich gives:
${\tilde g}_5 \, =\, 2 \pi $.

The parameter $ k$ is flavor dependent, representing the mass of the heavy quarks. The energy scale $1/z_0$ is taken as having the same value for charmonium and bottomonium, representing 
a flavor independent factor associated with just color interaction. The parameters used in ref. \cite{Braga:2015jca}  are:  
 \begin{equation}
  k_c = 1.2 GeV ; \,  k_b = 3.4 GeV ; \, 1/z_0 = 12.5 GeV,  
  \label{parameters}
  \end{equation}   
 \noindent where $ k_c$ and $k_b$ are the values of the constants $k$ used for charmonium and bottomonium, respectively.  
 Using these 3 parameters and the relations (\ref{new masses}) and (\ref{numerical decay})   the masses and decay constants of the states $1S, 2S, 3S, 4S$ of charmonium and bottomonium were estimated with
 an rms error of $30 \% $. 
 
In the next section we extend this model  to finite temperature and then, considering the same choice of  parameters of eq. (\ref{parameters}) we analyze  the behavior of  charmonium and bottomonium $S$ wave states in a thermal plasma.

\section{ Heavy Vector mesons at finite temperature }

Now we extend the zero temperature model of ref. \cite{Braga:2015jca}  to finite temperature.
It is important to mention that hadronic spectra at finite temperature have been studied in the context of  AdS/QCD soft wall model before,  for example,  in refs. \cite{Fujita:2009wc,Colangelo:2009ra,Miranda:2009uw,Fujita:2009ca,Mamani:2013ssa}.
 In particular \cite{Mamani:2013ssa} describes light vector mesons in the soft wall model. 
 However,  a complete analysis of the thermal spectral function for vector states of bottomonium and charmonium like the one performed in this article is not present in the literature.
 
 \subsection{Dual space and Hawking Page transition}
 
 Gauge string duality at finite temperature was discussed originally in refs.  \cite{Witten:1998qj,Witten:1998zw}. 
 Considering Euclidean signature and a compactified time coordinate, the geometry dual to
 a gauge theory at finite temperature is one of the two solutions of Einstein
 equations with constant negative curvature.  
 One of these solutions is the AdS black hole space that in Euclidean signature reads 
 \begin{equation}
 ds^2 \,\,= \,\, \frac{R^2}{z^2}  \,  \Big(    f(z) dt^2 + \frac{dz^2}{f(z) }  + d\vec{x}\cdot d\vec{x}  \Big)   \,,
 \label{EuclideanMetric}
\end{equation}
\noindent where the Schwarschild factor is $ f (z) = 1 - z^ 4/ z_h^ 4  $ and $z_h$ is the horizon position. 
 The other solution is the thermal AdS space,  that is just AdS space corresponding to $ f(z) = 1 $, with a compactified time. 

Following the work by Hawking and Page \cite{Hawking:1982dh},   one uses the semiclassical argument that there is  ``competition'' between the two solutions and the one with smaller Einstein Hilbert action will be stable at a given temperature. 
For the conformal gauge theory case (in a non compact space) the black hole is the stable solution 
for all temperatures\cite{Witten:1998zw}. So, the dual geometry is the black hole.  For  a non conformal gauge theory, as in the soft wall model case, the dual geometry has two different phases, as discussed in
refs. \cite{Herzog:2006ra,BallonBayona:2007vp}.  For temperatures above a critical value $T_c$ 
the black hole is stable while  for temperatures below $T_c$ the thermal  AdS is stable. The  so called  Hawking Page transition between spaces was interpreted in \cite{Witten:1998zw} as a transition in the dual gauge theory from a deconfined ($T > T_c$) to a confined phase ($T < T_c$).  

 In order to compare the action integrals of the black hole AdS and the thermal AdS we must take into account the fact that the periodicity of the time coordinate is related to the temperature. In our model the gauge theory is at $ z= z_0$ where the transverse part of the metric of the black hole is: 
 $$ ds^ 2 = \frac{R^ 2}{z_0^ 2}    \Big(    f(z_0) dt^2 +  d\vec{x}\cdot d\vec{x} \Big) \,.$$
The mapping of the supergravity theory to a gauge theory in flat  space must be performed with the rule that the gauge theory time has to be  $\tau = t \sqrt{ f(z_0) } $.  Since the period is the inverse of the gauge theory temperature: $ \tau \sim \tau + 1/T$ and the period of the black hole coordinate $t$ must be $\pi z_h$ to avoid a conical singularity at the horizon, one finds:
  
  \begin{equation} 
  T = \frac{1}{ \pi z_h \sqrt{ f (z_0) }}  = \frac{ 1}{ \pi z_h \sqrt{ 1 - \frac{z_0^4 }{z_h^ 4 }  \,}}\,.
  \label{temp}
  \end{equation}
 
 The action densities for the black hole AdS and thermal AdS in the soft wall model are calculated 
 in ref.  \cite{Herzog:2006ra}. The results of this article can be adapted  to the model considered here, where there is an UV cutoff, by replacing the minimum value of the coordinate $z$ that in ref. \cite{Herzog:2006ra} is just an UV regulator $ z= \epsilon $ by the (inverse of the) UV energy scale: $z = z_0$.
 Using also the relation between the horizon position and the temperature in eq. (\ref{temp}) 
 one gets:
 
 \begin{eqnarray}
 V_{th\, AdS} &=& \frac{4  R^ 3  }{g_5^2} \,\frac{1}{T}  \, \int_{z_0}^\infty dz \frac{ e^{-k^ 2 z^ 2}}{z^ 5} 
 \label{densities1} \\
 V_{BH \, AdS}  &=&   \frac{4 R^ 3  }{g_5^2} \, \frac{1}{T \sqrt{f(z_0) }} \int_{z_0}^{z_h} dz \frac{ e^{-k^ 2 z^ 2}}{z^ 5}  \,.
 \label{densities2}
 \end{eqnarray}
 
 The critical temperature, where the two actions densities are equal,  depends on the infrared parameter $k$ 
 of the soft wall background.  This parameter is flavor dependent. The formation of the plasma occurs when the lightest hadrons dissociate. So, we consider the confinement/deconfinement transition to be determined by the soft wall background that fits the masses of the $\rho$ vector mesons. In the present  model $\rho$ vector mesons can be described taking as in   ref. \cite{Braga:2015jca} $ 1/z_0 = 12.5 $ GeV 
and reproducing the calculation of the mass reviewed in section 2. One finds, using the parameter $ k $ = 0.388 GeV as in  \cite{Herzog:2006ra}, that the model with UV cut off leads to a mass
of 777.6 MeV for the $1 S $ state. 

The corresponding critical temperature is $ T_c = 191 $ MeV, the same  result of ref. \cite{Herzog:2006ra}.  In figure {\bf 1 } we  show the difference $ \Delta V = V_{BH \, AdS}  \,-\, V_{th\, AdS}\, $ between the action densities of eqs. (\ref{densities1}) and (\ref{densities2})  as a function of the temperature. The critical temperature corresponds to the point where the curve crosses the temperature axis.

\begin{figure}[t]
	\centering
		\includegraphics[width=0.75 \textwidth]{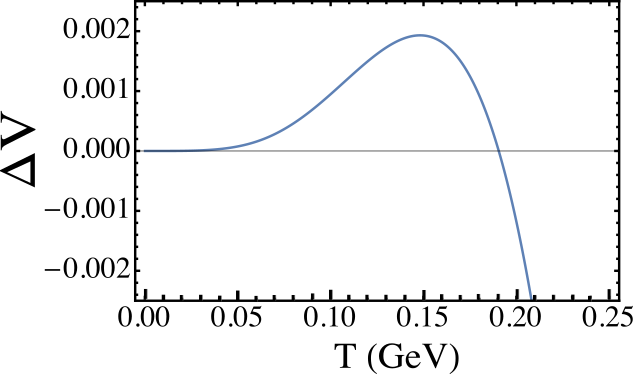}
	\caption{Difference between action densities of thermal AdS and black hole AdS as a function of the temperature for the model with UV cutoff. }
	\label{fig:figone}
\end{figure}

 \subsection{ Vector meson solutions in the black hole}
  
 As in the zero temperature case, we take a vector field $V_m = (V_\mu,V_z)\,$ ($\mu = 0,1,2,3$)  described by an action integral with the general form of eq. (\ref{vectorfieldactionzeroptemp}) and soft wall background $\Phi = k^2z^2   $ . But   for describing the thermal spectra one considers the geometry as
the Minkowski version of the black hole metric (\ref{EuclideanMetric}): 
\begin{equation}
 ds^2 \,\,= \,\, \frac{R^2}{z^2}  \,  \Big(  -  f(z) dt^2 + \frac{dz^2}{f(z) }  + d\vec{x}\cdot d\vec{x}  \Big)   \,.
 \label{MinkoviskyMetric}
\end{equation}
where again: $ f (z) = 1 - z^ 4/ z_h^ 4  $ and  the gauge theory temperature is related to the horizon position by eq. (\ref{temp}). 
It is important to note that this black hole geometry will be stable only  for temperatures $T > T_c$. 
We will  calculate the thermal spectral functions using  this black hole  metric for all temperatures
with the interpretation that for  $T <  T_c$ it represents  a super-cooled (unstable)  phase. 

As in the zero temperature case,  we choose the gauge $V_z = 0 $ and assume  $ V^0_{\mu}(x) =
\lim_{z\to z_0} V_\mu (x,z) $ to be the sources of the correlation functions of  $  J^\mu (x) \,$.  
Now, with the radial AdS coordinate defined in the region: $ z_0 \le z \le z_h $, the on shell action takes the form: 
  \begin{equation}
I_{on \, shell }\,=\, - \frac{1}{2 { g}_5^2}  \, \int d^4x \,\,\,\Big[ e^{- k^2 z^ 2  } \sqrt{-g}  \, g^{zz} g^{\mu \nu}   V_\mu \partial_z V_\nu  \Big]
{\Big   \vert}_{_{ \! z \to z_0 }}^{_{ \! z \to z_h }}
 \,.
\label{onshellaction2}
\end{equation}
 The imaginary part of the on shell action should generate holographically the thermal spectral function.
 However, it was pointed out in ref. \cite{Son:2002sd} that for an action like (\ref{onshellaction2}) 
  the imaginary part is  $z $ independent. So the contributions from the two integration limits 
cancel each other out.   
This problem can be solved following again  \cite{Son:2002sd} and using the additional prescription that only the boundary $z= z_0$  is considered.   In other words, one takes eq. (\ref{onshellaction2}) with only the lower integration limit. For an interesting discussion on the interpretation of the prescription for calculating the retarded Green's function, see \cite{Iqbal:2008by}.

The procedure to find the retarded Green's function involves fourier transforming the fields and decomposing the momentum space fields as it was done in the finite temperature case in eq. (\ref{Bulktoboundary}):
$ V_\mu (q,z) \,=\, v (q,z) V^0_\mu ( q ) \,$. The on shell action takes the form
  \begin{equation}
I_{on \, shell }\,=\,   \, \int d^4q \,\,\,\Big[  {V^ 0_\mu}^\ast ( q ) {\cal F}^{\mu\nu} (z,p)  V^0_\nu ( q )
  \Big]_{_{ \! z \to z_0 }}
 \,,
\label{onshellaction3}
\end{equation}
where 
\begin{equation}
 {\cal F^{\mu \nu} } (z,q)  \,=\,   \frac{1}{2 g_5^2}   e^{- k^2 z^ 2  }  \sqrt{-g}  \, g^{zz}   g^{\mu\nu} 
v^\ast  ( q,z)  \partial_z v (q,z) ]\,.
\end{equation}

The corresponding retarded Green's function is:
\begin{equation}
 G_R^{\mu\nu}  (q) \,=\, {\cal F}^{\mu\nu}  (z=z_0 , q)  \,,
\end{equation}   
and the spectral function is the imaginary part of the retarded Green's function: 
\begin{equation}
 \rho^{\mu\nu}  (q) \, =\, -  {\cal I}m  \{ G_R^{\mu\nu}  (q) \}   \,.
\end{equation}
 The bulk to boundary propagators $ v (q,z) $ are solutions of the equations of motion. These equations  have different forms for the temporal $V_0$ and spatial $V_i$ components of the vector field. For the case of a plane wave solution with momentum $q^\mu = (\omega , \vec q )  $ they are:
\begin{eqnarray}
\partial_z \Big( \frac{ e^{-k^ 2 z^ 2 }  } { z}  \partial_z V_0 (q,z) \Big) &-&
 \frac{ e^{-k^ 2 z^ 2 } }{ z f(z) } \Big(   \frac{\omega^ 2 }{ f(z) }  -\vert {\vec q}\vert^2 \Big)  
   V_0 (q,z) \,=\, 0  \cr \cr
      \partial_z \Big( \frac{ e^{-k^ 2 z^ 2 } f(z)  } { z}  \partial_z V_i (q,z) \Big) &+&
 \frac{ e^{-k^ 2 z^ 2 } }{ z } \Big(   \frac{\omega^ 2 }{ f(z) }  - \vert {\vec q}\vert^2 \Big)  
   V_i (q,z) \,=\, 0  \,.
       \label{EOMfiniteT}
\end{eqnarray}

It is convenient \cite{Grigoryan:2010pj}  to choose the momentum  $q^\mu = (\omega , \vec 0 ) $ where the transversality of the current $ q^\mu J_\mu = 0 $ translates  into the vanishing of the temporal component $J_0$.
Then we just need to solve the equation for the spatial component:
$ V_i (\omega ,z) \,=\, v (\omega ,z) V^0_i ( \omega ) $. In this case $ v (\omega ,z) $ satisfies the equation:

\begin{equation}
 \partial_z \Big( \frac{ e^{-k^ 2 z^ 2 } f(z)  } { z}  \partial_z v (\omega ,z)  \Big) +
 \frac{ e^{-k^ 2 z^ 2 } }{ z }   \frac{\omega^ 2 }{ f(z) }    
   v (\omega ,z) \,=\, 0 \,.
   \label{eqmotionbtb}
 \end{equation}

The bulk to boundary propagator has to satisfy two boundary conditions. One is 
\begin{equation}
v (\omega, z=z_0) = 1\,,
\label{conditionbtb}
\end{equation}
that was present in the zero temperature case and implies that the field components work as the sources of the correlation functions at $z= z_0$. The other is the condition that the solution behaves as an incoming wave in the near horizon limit
$ z \to z_h$. The absence of outgoing solutions represents the absorption by the black hole horizon. 
In order to implement this condition one can use the Regge-Wheeler tortoise coordinate that makes it explicit the decomposition of the solutions of the equations of motion in incoming plus outgoing solutions.
One introduces the coordinate  $ r_\ast $ such that $ \partial_{r_\ast} = - f(z) \partial_z $ that implies 
\begin{equation}
 r_\ast  = \frac{1}{2} z_h \Big[ -\tan^{-1} \left( \frac{z}{z_h} \right)  +
  \frac{1}{2}  \ln \left(  \frac{z_h - z}{z_h + z} \right)  \Big]\,,
 \end{equation} 
 in the interval  $ z \le z_h $ where $z$ is defined.

 Performing a Bogoliubov transformation $ v (\omega , z) = e^{B/2} \psi  (\omega , z) $ with $ e^B = z e^{k^ 2 z^ 2}  $
 one finds that the equation of motion (\ref{EOMfiniteT})  takes the form
 
 $$ \partial^2_{r_\ast}  \psi + \omega^2 \psi = U \psi \,, $$
 
 \noindent where the potential
 \begin{equation} 
 U (z) \,=\, \left( 1 - \frac{z^ 4}{z_h^4} \right) \left[ \left(  k^4 z^2 + \frac{3}{4z^ 2} \right)   \left( 1 - \frac{z^ 4}{z_h^4} \right) +   2 z^2 \frac{(1 + 2 k^2 z^2 )}{z_h^4} \right] 
 \end{equation} 
vanishes at the horizon. Thus, the function $\psi$ has the asymptotic near horizon solutions $\psi_{in/out} = e^{ \mp i \omega  r_\ast} $ representing incoming and outgoing waves respectively.

 Expanding the incoming wave solution near the horizon as 
 \begin{equation}
  \psi_{in} =  e^{ - i \omega  r_\ast} \left[ 1 + a_1 (z - z_h) + a_2 (z - z_h )^2 + ... \right] \,,
 \end{equation}
 and inserting in the equation of motion one finds the relevant coefficient: 
 \begin{equation}
 a_1 =   \frac{ 1 + 2k^2 z_h^2 }{ z_h (  i \omega z_h - 2 ) } \,.
 \label{a1}
 \end{equation}
 In order to implement the incoming wave condition we write the bulk to boundary propagator as
 \begin{equation}
v (\omega, z) \,=\,e^{ - i \omega  r_\ast} \, F(\omega,z) \,,
\label{vdecomp}
\end{equation}
so that the function $F$  takes the form:
\begin{equation}
  F(\omega,z) = \sqrt{z} e^{\frac{k^2z^2}{2}}  \left[ 1 + a_1 (z - z_h) + a_2 (z - z_h )^2 + ... \right]
\label{BCF}
\end{equation} 
and the derivative of $F $ at the horizon is obtained from this expansion and the expression for $a_1$ in eq. (\ref{a1}).

Finally, the spectral function for spatial components $\rho^{ii} $ with the choice of momentum $q^\mu = (\omega , \vec 0 ) $ and written in terms of $F $ takes the form (omitting the indices $ i i $)  
\begin{equation}
 \rho (w) \,=\,   \frac{w } {2  {\tilde g}_5^2}  \, \frac{ e^{- k^2 z_h^ 2 }    } { z_h}  \,
\vert F( \omega , z_h) \vert^ 2  \,,
\label{spectralf}
\end{equation}
where we defined the dimensionless coupling $ {\tilde g}_5^2 =  g_5^2 /R $, as in the zero temperature case.   This is the object that will describe the thermal behavior of the heavy vector mesons. In the next section we
 present the results of the numerical calculations of $\rho$.

\section{  Spectral functions for charmonium and bottomonium S-wave states}

\begin{figure}[t]
	\centering
		\includegraphics[width=1 \textwidth]{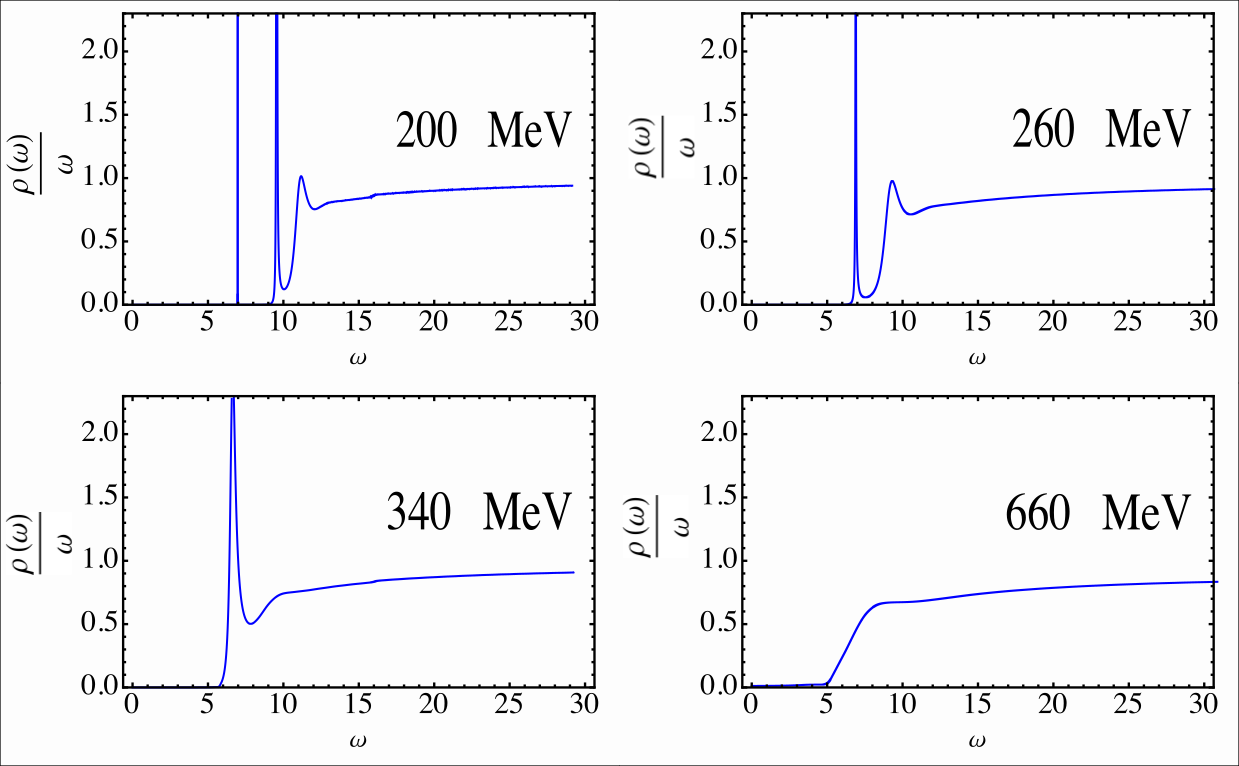}
	\caption{Bottomonium melting process starting at 200 MeV with three states 1S, 2S and 3S at left upper panel. Each  panel shows the melting temperature for these states. }
	\label{fig:figtwo}
\end{figure}

\begin{figure}[t]
	\centering
		\includegraphics[width=1 \textwidth]{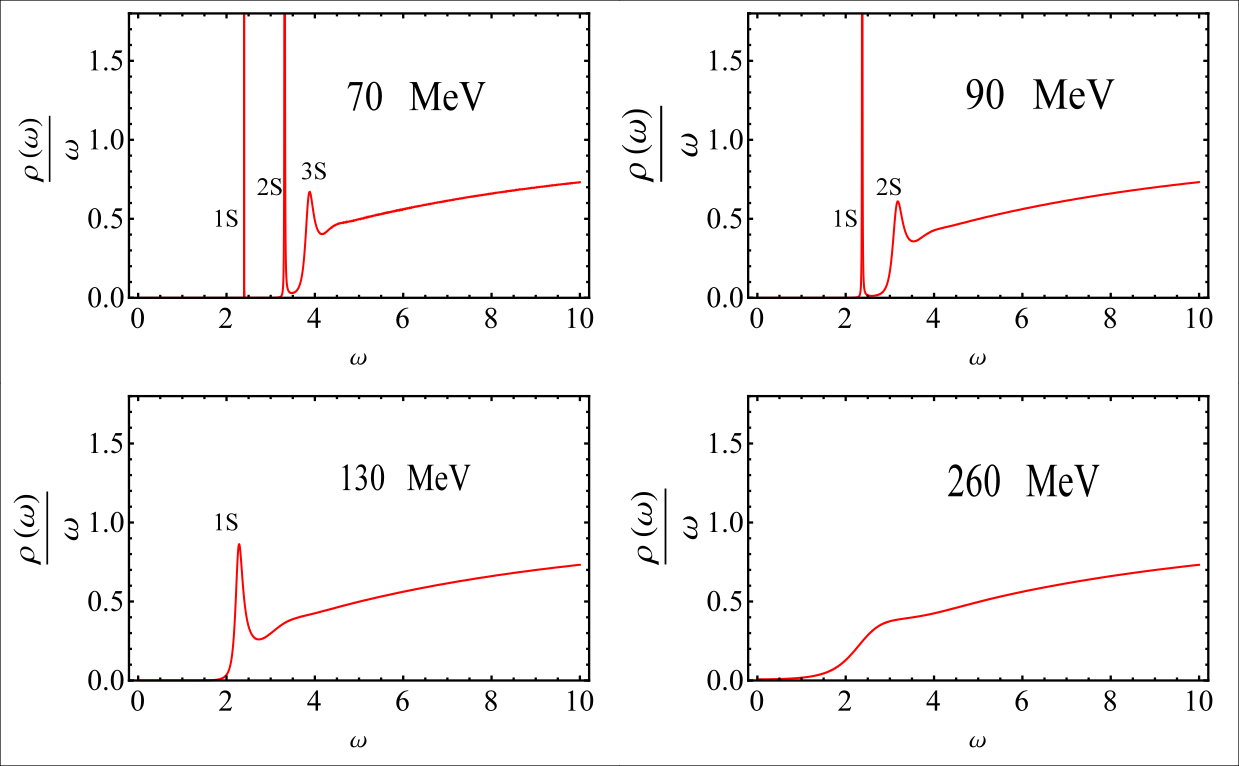}
	\caption{Charmonium melting process starting with a  temperature of 70 MeV with three initial states 1S, 2S and 3S at left upper panel. Each of the three remaining panels shows the melting temperature of these states.}
	\label{fig:figthree}
\end{figure}

 We solved numerically equation (\ref{eqmotionbtb}) for the bulk to boundary propagator $v (\omega, z) $, written in terms of the function $ F$ as in eq. (\ref{vdecomp}),  with the boundary conditions described in the previous section. 
  The parameters used are the zero temperature ones, from ref. \cite{Braga:2015jca}, namely a flavor independent UV cutoff $  1/z_0 = 12.5 $  GeV and flavor dependent soft wall parameters with values  $ k_c = 1.2 $ GeV for charmonium and $  k_b = 3.4 $ GeV  for  bottomonium S-wave states. 
 
 The spectral function (\ref{spectralf})  was calculated for different temperatures. An important non trivial fact emerged from the  analysis of the large frequency asymptotic behavior.  It is well know that when one calculates the spectral function from correlators at the conformal boundary $z \to 0$, the spectral function in the limit 
 $\omega \to \infty$ grows up as $\rho \sim  \omega^2 $. This results comes from  conformal invariance and dimensional analyisis (see for example ref. \cite{Myers:2007we}). 
 
 In the present case we do not calculate the correlators at the $z\to 0$ conformal limit. There is an extra dimensionfull quantity, the position  $z_0$,  that appears in the calculation of the spectral function.  So, the argument of simple dimensional analysis does not hold  in the same way here. 
The numerical results obtained show a behavior that is different from the conformal case. For large frequencies the spectral function   grows linearly with the frequency: $ \rho \sim \omega $.  We present in  appendix {\bf B} an analysis of this behavior. We show there that if 
 in the present  model one takes the limit of $z_0 \to 0 $ one finds spectral functions growing with $\omega^2$, as expected in the conformal case. But for the finite value of $ z_0$ explored here they grow with $\omega$ for large $\omega$.  So, we analyzed the behavior of the relevant (normalized) quantity: 
 $$  \frac{ \rho(\omega ) }{\omega } \,.$$
 
 We show in figure {\bf 2} the spectral functions for the bottomonium vector states at four illustrative temperatures. In these plots one can clearly see the following situation: 
 \begin{itemize}
 \item[] $\bullet$  
  at $T = 200$ MeV three peaks   corresponding to $ 1S$, $2S$ and $3S$ states; 
 \item[] $\bullet$    at $T = 260$ MeV two peaks corresponding to the melting of the $3S$  state;
  \item[] $\bullet$ 
   one peak at $T = 340 $ MeV where only the $1S$ states survives and
   \item[] $\bullet$  at $T = 660 $ MeV  the complete melting of the states. 
 \end{itemize} 
 We present in the appendix {\bf A}  a more detailed picture of the melting process by showing more plots that illustrate the temperature evolution of the spectral function. From this analysis one can infer that the states $1S, 2S $ and $3S$ melt at different temperatures, as expected. In particular, the $1S$ states survives at temperatures much larger than the critical temperature. 
 The complete disappearance of the $1 S$ states happens at $ T \sim 600 $ MeV, corresponding to $ T/T_c \sim 3.2  $. For the $ 2S$ state there will be no trace of the peak for temperatures above $ T \sim 360 $ Mev, corresponding to $ T/T_c \sim 1.9 $ while for the $3 S$ states the total melting happens at $ T \sim 220 $ MeV, that means $T/T_c \sim  1.2  $.

  Then figure {\bf 3 } shows the spectral functions for the charmonium vector states at  four different  temperatures
 that illustrate the melting process. More details for the thermal evolution of charmonium states are shown in appendix {\bf A} . One can clearly see the change  from the case with three well defined peaks corresponding to the states $1S, 2S$ and $3S$  to the case where there is no well defined quasi particle state. An important difference with respect to the bottomonium case is that the melting process occurs at temperatures below $T_c$. At the critical temperature  there is only a very small peak of the state $1S$, so
 one can interpret this situation as meaning that the charmonium states $ 2S$ and $3S$ do not survive in the deconfined plasma phase, while there could be  some trace of the $1 S$ state up to temperatures of $ 1.2 T_c$.

 The present results for bottomonium states are consistent with the ones obtained using lattice QCD in \cite{Suzuki:2012ze}. This article predicts a lower bound for  the melting temperature of the 1$S$ state of
 $ 2.3 T_c$. They are also consistent with the lattice results of \cite{Aarts:2011sm} where the temperature range between $0.4 T_c$ and $2.1 T_c$ was analyzed and the $1 S$ state survives for higher temperatures whereas the higher excitations melt around $ 1.4 T_c$. 
 It is interesting to mention that experiments show that in Au + Au collisions  with center of mass energy of 200 GeV  the bottomonium states $ 2 S $ and $ 3 S$ are completely  suppressed \cite{Adare:2014hje}.
 
 Using a potential model, ref. \cite{Mocsy:2007jz} finds that the excited states of charmonium melt below $ T_c$ while the $1S$ state melts at $ 1.2 T_c$,  that is consistent with our results, taking into account the error that will be discussed in the next section. 
 
 The results obtained here are also consistent with the anlysis of the  thermal behavior of quarkonium states using QCD sum rules developed in refs. \cite{Dominguez:2009mk,Dominguez:2010ve,Dominguez:2013fca} regarding the survival of quarkonium states above the critical temperature.

\section{  Conclusions  }

It is shown in this paper that a consistent picture for the thermal behavior of $S-$wave states of bottomonium and charmonium emerges from a finite temperature version of the model for heavy vector mesons masses and decay constants proposed in ref. \cite{Braga:2015jca}. The spectral functions obtained numerically for bottomonium and charmonium states exihibit  clear peaks for the states $1S, 2S $ and $3S$ at low temperatures.  As the temperature increases, the peaks spread and disapear, with the expected result that highly excited states melt in the thermal medium (plasma) at lower temperatures. 

One point that must be remarked is that the model of reference \cite{Braga:2015jca} presents a rms error of $ 30 \% $ when one fits the decay constants and masses of the  four initial $S$ wave states of charmonium and bottomonium. So, one should not consider our numerical results for the melting temperatures of the states with a precision larger than that. We mean, our (rougth) estimate for the error
in the melting temperatures is of the order of $ 30 \%$. 

Even with this error, one can infer that the model predicts a very distinct behavior for bottomonium and charmonium states.  This could be an iteresting tool to investigate not only the formation of quark gluon plasma but also the temperature os the thermal medium. The strong supression of charmonium states with a low supression of bottomonium states would indicate temperatures not much larger then the critical one. On the other hand, an eventual observation of  supression of bottomonium $S$ wave states could indicate the formation of plamas at higher temperatures. 

One question that could be asked is if one could find more acurate etimates for the melting temperatures using holography.  With more accurate results one could be more confident in analysing the temperture of the plasma from the relative supression of the different states. 
An alternative model for calculating masses of heavy vector mesons was recently proposed in ref.  
\cite{Braga:2015lck} . In this reference the masses of charmonium and bottomonium states are estimated with an rms error of $2.0 \%$.  It would be nice to formulate a finite temperature version of this model also, in order to compare the thermal behavior with the one found here. There is however an obstruction to this task. The incoming wave condition that has to be used for the field that describes a vector meson at finite temperature is apparently inconsitent with the zero temperature limit of the incoming wave condition at the 
black hole horizon. More precisely,  at any finite temperature, the incoming wave condition implies that the derivative of the bulk to boundary propagator is infinite at the horizon. In the limit of zero temperature this would mean that the derivative should be infinite at $ z \to \infty$. In contrast, in the model of ref. \cite{Braga:2015lck}  there is the boundary condition that the derivative of the bulk to boundary propagator is zero at $z\to \infty$. We leave for a future work the non trivial task of finding a consistent finite temperature for this model.  

\vskip1cm

\noindent {\bf Acknowledgments:}  We thank  Jorge Noronha for suggesting that we study the finite temperature version of ref. \cite{Braga:2015jca} and for very important comments about a draft  of the manuscript. 
N.B. and S.D. are partially supported by CNPq and M.A.M. is supported by Vicerrectoria de Investigaciones de La Universidad
de los Andes.

\begin{figure}[t]
    	\centering
    	    \includegraphics[width=1 \textwidth]{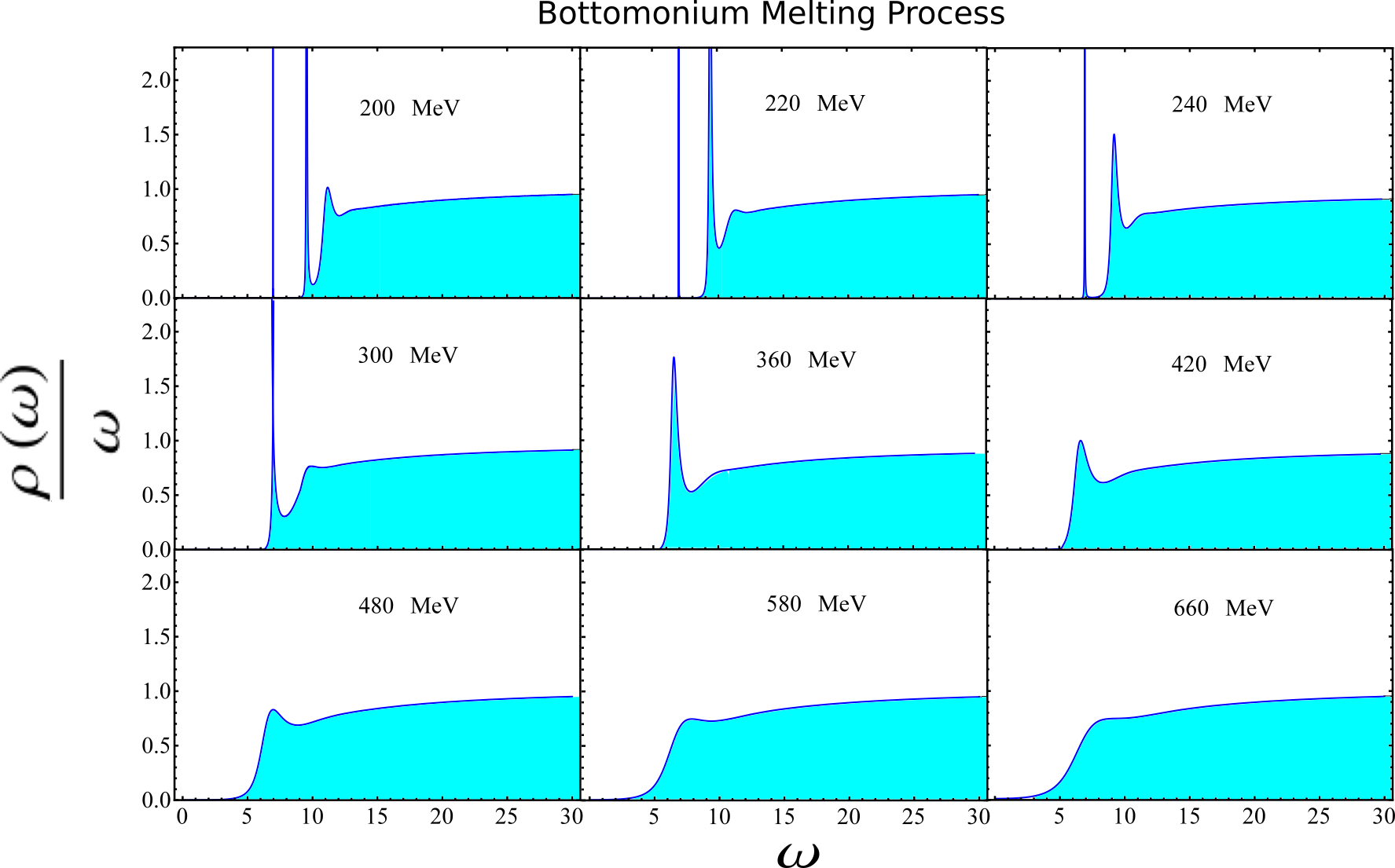}
    	    \caption{Complete Bottomonium melting process starting at 200 MeV with three states: 1S, 2S and 3S. The states 3S and 2S melt at temperatures near to 220 MeV and 300 MeV respectively. The 1S state melts at temperature near to 580 MeV. }
    	    \label{fig:figfour}	    
\end{figure}

\begin{figure}[t]
    \centering
        \includegraphics[width=1 \textwidth]{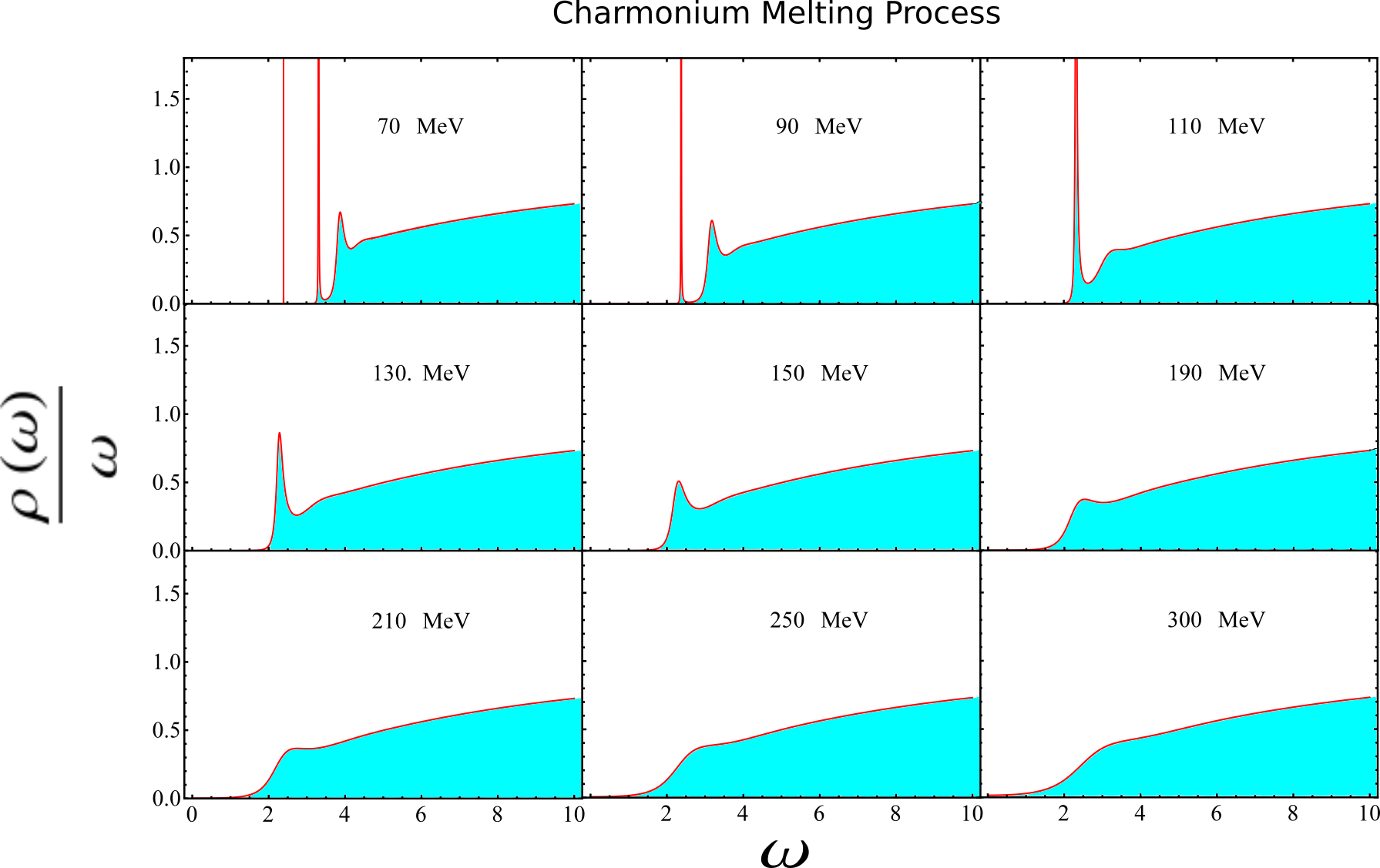}
        \caption{Complete Charmnonium melting process starting at $T=70$ MeV, where we have three states: 1S, 2S and 3S. At a temperature about 90 MeV the 3S  state melts. The 2S state melts down near to 110 MeV and finally,the 1S state melts at about 250 MeV.}
        \label{fig:figfive}
\end{figure}

 \newpage   

\appendix

\bigskip

\noindent {\large \bf Appendix }

\section{  Temperature dependence of the spectral functions}

In order to present a more detailed view of the bottomonium melting process, we show in figure {\bf 4} the thermal spectral function for nine different representative temperatures.
  At 200 MeV, we have three defined vector states $\Upsilon  (1S)$, $\Upsilon^\prime (2S)$ and $\Upsilon^{\prime\prime} (3S)$. At  220 MeV, one can see that the 3S state disappears. 
  So, the 3S melting temperature in this model is between 200 MeV and 220 MeV. 
  Then the 2S peak disappears near  300 MeV (left middle panel). 
  Finally, near 580 MeV one observes the 1S melting. Lattice calculations \cite{Adare:2014hje} 
  show that the $\Upsilon$ (1S) melting temperature lies inside an interval  350 MeV--612 MeV, while for the $\Upsilon$ (2S) and $\Upsilon$ (3S) the melting temperatures are in a 200 MeV-- 300 MeV region. Our results are consistent  with these calculations.

Figure  {\bf 5 } shows the behavior of charmonium spectral function. The panels correspond to temperatures varying in steps of 20 MeV. Starting at the upper left panel, at the temperature of 70 MeV there are three peaks corresponding to  $J/\Psi$, $\Psi'$  and $\Psi''$. At higher temperatures one observes the melting starting by the heavier states.
At $T =$ 90 MeV, the 3S state melts. Then at temperatures about  110 MeV  the 2S melts. 
Then at $ T = 250 $ MeV the  1$S$ peak has virtually disappeared.

It is important to take into account the fact that for temperatures below $T_c$ the black hole phase is unstable due to Hawking Page transition, as explained in section {\bf 3}. So, the transitions described in the plots of  lower temperatures could be absent if the plasma phase is not formed and the medium is confined.
So, the thermal spectrum is more reliable for  $T > T_c = 191$ MeV.

\begin{figure}[h]
	\centering
		\includegraphics[width=1.0 \textwidth]{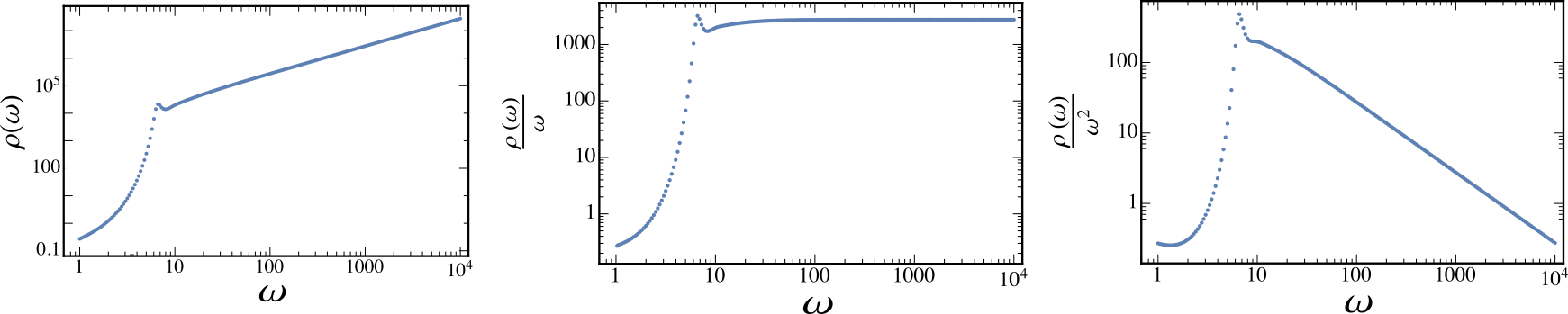}
	\caption{Spectral function for our model: $z_0=1/12.5GeV$. The second plot ($\rho/\omega$) is constant for large $\omega$. }
	\label{fig:figone}
\end{figure}

 \begin{figure}[h]
	\centering
		\includegraphics[width=1.0 \textwidth]{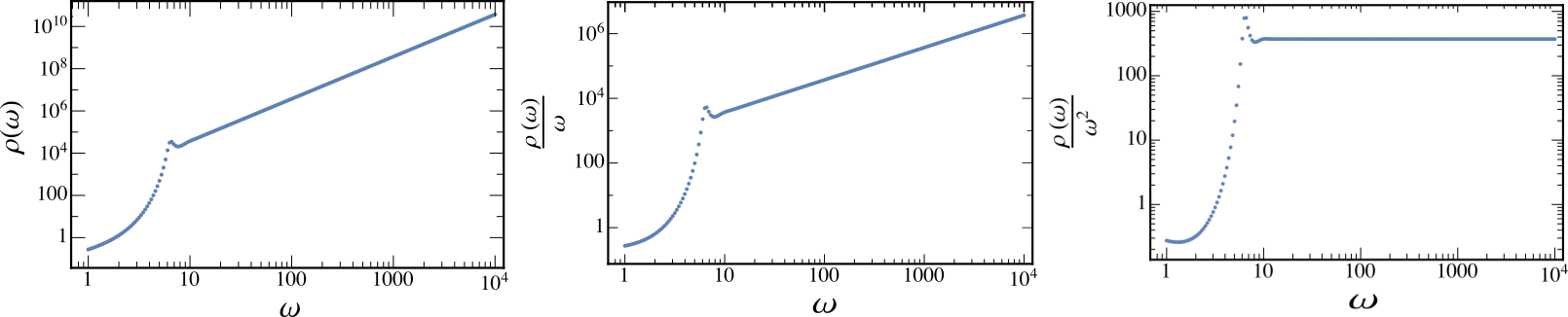}
	\caption{Spectral function near the conformal boundary: $z_0=10^{-6}GeV^{-1}$. The  third plot shows 
	that $\rho/\omega^2$ is constant for large $\omega$.}
	\label{fig:figone}
\end{figure}

\section{ High energy behaviour of the spectral functions}

At high frequencies,  the spectral functions studied  in this article show a non trivial behavior.
The holographic model presented in section {\bf 3 } and extended to finite temperature is section {\bf 4}, with two point correlation functions calculated at a finite position $z = z_0=1/(12.5 GeV)$ of AdS space, leads to spetral functions   $\rho(\omega) \propto \omega$ in the 
limit of large $\omega$. 
This result contrasts with the situation when gauge theories correlators are calculated at $ z=0$ and  conformal symmetry is manifest implying:  $\rho(\omega)\,\propto \, \omega^2$.

In order to  display the effect of the $z_0$ parameter in the assimptotic behaviour of spectral functions,  we plot in logarithm scale in separate panels 
$\rho(\omega)\,,~\rho(\omega)/\omega\,$ and $ ~\rho(\omega)/\omega^2$ for frequencies up to $10^4 GeV$ using two  different choices of $z_0$. 
Since we are interested only in the role played  by the parameter  $z_0$, we fix the temperature and the dilaton constant $k$ in all plots to the values: $T=400 MeV,~ k=3.4 GeV$.

In figure {\bf 6} we choose the parameter:  $z_0=1/(12.5 GeV)$ that was used in the present article. One clearly see in the second panel that  $\rho/\omega$ reaches a constant value  for  $\omega \gtrsim 50 GeV$. As a check, the first panel shows the increase of  $\rho $ and the third the decrease of
 $\rho/\omega^2$ for large $\omega$.

Then, as check, one can take the  limit where the present model should recovers the usual soft wall  case, namely, a very small $z_0$. We show in figure {\bf 7} the situation at $z_0 = 10^{-6}GeV^{-1}$. 
Consistently, one observes that in this case where $z_0$ approximately ceases to be a parameter of the model, the ultraviolet behavior of the spectral function changes to  $\rho \propto \omega^2$.

 \end{document}